\documentclass[a4paper]{jpconf}
\usepackage{graphicx}
\newcommand{\be}{\begin{equation}}
\newcommand{\ee}{\end{equation}}

\newcommand{\bea}{\begin{eqnarray}}
\newcommand{\eea}{\end{eqnarray}}
\newcommand{\bean}{\begin{eqnarray*}}
\newcommand{\eean}{\end{eqnarray*}}
\begin{document}
\title{Geometrical Expression of the Angular Resolution of a Network of Gravitational-Wave
  Detectors and Improved Localization Methods} 

\author{Linqing Wen$^{1,2,3}$, Xilong Fan$^4$, and  Yanbei Chen$^{1,2}$}

\address{$^1$ Max Planck Institut f\"ur Gravitationsphysik,
  Albert-Einstein-Institut, Am M\"uhlenberg 1, D-14476, Golm,
  Germany}
\address{$^2$ Division of Physics, Mathematics, and Astronomy, Caltech, Pasadena, CA 91125, USA} 
\address{$^3$ School of Physics, University of Western Australia, Crawley, WA 6009, Australia}
\address{$^4$ Department of Astronomy, Beijing Normal University,
Beijing, China}

\ead{lwen@aei.mpg.de}

\begin{abstract}
We report for the first time a method-independent geometrical expression for the angular resolution of an arbitrary 
network of interferometric gravitational wave (GW) detectors when the
arrival-time of a GW is unknown.  We discuss the implications of our results on how to improve angular resolutions of a GW network and on improvements of localization methods.  An example of an improvement to the null-stream localization method for GWs of unknown waveforms is demonstrated.
\end{abstract}

\section{Introduction}
Several types of astrophysical sources are expected to be detectable both in gravitational waves (GWs) and in electromagnetic waves.  Coincidence detections of these sources are of significant astronomical interest \cite{wen07m,hughes06}.  A clear understanding of the angular resolution of an array of multiple GW detectors is vital to localizations of GW sources and to coincident detections. 

%It is known that angular direction of a GW source can be obtained by measuring differences in arrival-times of the wavefront between different detectors.   
A standard approach to measure how well we can localize a source 
is to calculate  the Fisher information
matrix where method-independent lower-bounds on statistical
errors of estimated parameters can be obtained. Numerical results have been calculated by many authors for angular resolutions of both the ground-based and the future space GW detector \cite{krolak94,cutler98,Pai01,leor04}.
Explicit analytical expressions for the angular
resolution of a network of GW detectors have been rare in the literature.   We found two approximate formulae for a 3-detector network which are summarized in \cite{sylvestre03}.  One is from private communication of Thorne
(cited in Ref.~\cite{tinto89}).  The other is based on normalized 
numerical results for a 3-detector network for detections of GWs from neutron
star-neutron star coalescence using the coherent
approach~\cite{Pai01}. A general expression for an arbitrary network
of GW detectors have not been seen.  

%We will summarize our result of such a general expression from our on-going work \cite{wen07a}. 

Localization of GW sources of unknown waveforms can obtained by the so-called ``null-stream'' method \cite{tinto89,wen05}.  GWs are known to have only two polarizations.  The
response of an interferometric GW detector is a
linear combination of the two wave polarizations.  Therefore if we
have data from more than two detectors, we can linearly combine the 
data to cancel out the GW signal. The resulting data streams are
called the ``null-streams'' as they have null-responses to GW signals.
Localization of a GW source can be achieved by searching for sky
directions where the constructed null-stream is statistically
``null'' \cite{tinto89,wen05}.  There are also semi-null streams \cite{wen07} where in linearly combined data, signals are not
exactly canceled out but are significantly reduced. 
We propose that localization be further improved by including information from these semi-null streams.

In this report, we summarize results of our on-going research work concerning (1) the angular resolution of an arbitrary network of interferometric GW detectors \cite{wen07a} and (2) localization methods for GWs of unknown waveforms \cite{wen07b}.  An explicit geometrical expression for the angular resolution of an arbitrary network of GW detectors is presented for the first time.  An improved localization method using null-streams combined with semi-null streams is demonstrated and compared to that of a straightforward null-stream-only method.

\section{Mathematical Preliminaries}
\label{principle}
Suppose we have a network of $N_d$ gravitational-wave detectors, each with spatial size much shorter than the GW wavelength,  the observed strain of an incoming GW by the individual detector $I$ is then a linear combination of the two wave  polarizations in the transverse traceless (TT) gauge,
\begin{equation}
d_I(t_0+\tau_I+t)=f^{+}_Ih_{+}(t)+f^{\times}_Ih_{\times}(t),\quad 0<t<T\,,
\end{equation}
where  $t_0$ is the arrival time of the wave at the coordinate origin, $\tau_{I}$ is the wave
travel time from the origin to the $I$-th  detector, $T$ is the signal duration, $t  \in [0,T]$ is
the time label of the wave.  The
quantities  $f^{+}$ and $f^{\times}$ are the detector's {\it antenna
  beam pattern} functions \cite{krolak98} for the two wave
polarizations ($h_+$, $h_{\times}$).  They depend on the relative
orientation between the detector configuration and the frame in which
the wave polarizations are defined (which is in turn related to the propagation direction $\hat{\mathbf{n}}$). 

If we assume signal duration to be short enough such
that the motion of the detector array is unimportant, then in the
frequency domain, and in matrix notation, we can write time-delay-shifted  responses of all
detectors as
\begin{equation}
\label{dorig}
\mathbf{d}(\Omega)  =\mathbf{A} \mathbf{h}(\Omega),
\end{equation}
where $\Omega$ is the angular frequency. The antenna pattern $\mathbf{A}$ is an $N_d \times 2$ constant matrix, 
\begin{equation}
\mathbf{A} =
\left[
\begin{array}{cc}
f^+_1(\hat{\mathbf{n}}) & f^\times_1(\hat{\mathbf{n}}) \\
\vdots & \vdots \\
f^+_{N_d}(\hat{\mathbf{n}}) & f^\times_{N_d}(\hat{\mathbf{n}})
\end{array}
\right], 
\end{equation}
and $\mathbf{h}(\Omega)$ is a 2-dimensional vector function,
\begin{equation}
\mathbf{h}(\Omega)
=
\left[
\begin{array}{c}
h_+(\Omega) \\
h_\times(\Omega)
\end{array}
\right]\,.
\end{equation}

We denote $d_I$ as the data from the $I$-th GW detector and the corresponding noise spectral density is $S_I$,  we define a whitened data set of 
\begin{equation}
\hat d_I (\Omega) = S_I^{-\frac{1}{2}} (\Omega) d_I(\Omega)\,.\quad
\end{equation}
Note that $\hat{\mathbf{d}}(\Omega)$ corresponds to the whitened data set at each frequency. Correspondingly, we
denote $\mathbf{\hat{A}}$ as a $N_d \times 2$  response matrix
weighted by noise,
\begin{equation}
\hat{\mathbf{A}} (\Omega) \equiv 
\left[
\begin{array}{cc}
\frac{f^+_1(\hat{\mathbf{n}})}{\sqrt{S_1(\Omega)}} & \frac{f^\times_1(\hat{\mathbf{n}})}{{\sqrt{S_1(\Omega)}}} \\
\vdots & \vdots \\
\frac{f^+_{N_d}(\hat{\mathbf{n}})}{\sqrt{S_{N_d}(\Omega)}} & \frac{f^\times_{N_d}(\hat{\mathbf{n}})}{\sqrt{S_{N_d}(\Omega)}}
\end{array}
\right],
\label{A_hat}
\end{equation}
so that we have $\hat{\mathbf {d}} = \hat{\mathbf{A}} \mathbf{h}$. For
simplicity, we keep the $\Omega$-dependence in the notation only when
it is necessary for clarity. 

\section{Geometrical Expression of Angular Resolution}\label{null stream}
The angular resolutions are calculated by  applying the Fisher information matrix to obtain method-independent lower limits on the statistical errors in estimating the direction of a GW source.  The limits are for unbiased estimators and Gaussian noise (for cautions in using these limits,
see \cite{michele07}).  The covariance matrix of the ``best estimated''  angular direction of a GW source can be obtained from the corresponding sub-matrix
of the inverse of the Fisher matrix for all unknown parameters.  We show in \cite{wen07a} that in case the initial arriving time $t_0$
of the wave is unknown, the lower bounds of one-sigma error area of angular parameters estimated using data from an
arbitrary network of GW detectors  can be written in a compact
geometrical form. Here we only summarize the result without showing
derivations.  We present also only cases for short-duration GWs where antenna beam patterns of GW detectors are treated as constant. Similar expressions for continuous GWs for ground-based detectors and for the space GW detector LISA can be found in \cite{wen07a}. 

We have defined the
error solid angle to be twice the area of
the 1-$\sigma$ error ellipse (measured in {\it srad}) in
angular parameters of $\theta$ (latitude-like)  and $\phi$ (longitude-like), 
\begin{equation}
\Delta\Omega = 2\pi |\cos{\theta}| \sqrt{\langle \Delta \theta^2\rangle \langle\Delta
  \phi^2\rangle - \langle\Delta \theta \Delta \phi\rangle^2}\,,
\end{equation} 
we have found that for an arbitrary network of GW detectors, 
\begin{eqnarray}
&&\Delta\Omega = \frac{4\sqrt{2}\pi c^2}{\sqrt{\sum_{J,K,L,M} \Delta_{JK}\Delta_{LM} |(\mathbf{r}_{KJ} \times \mathbf{r}_{ML})\cdot \hat{\mathbf{n}}|^2}},
\label{T_geo}
\end{eqnarray} 
where $\mathbf{r}_{KJ}$ is the displacement vector from detector K to
detector J. 

For the worst-case scenario where nothing is known about the initial
arrival time $t_0$ or the waveform of a GW, we found that
\begin{eqnarray}
\Delta_{KJ} &=&\frac{1}{\pi} \int_{-\infty}^{+\infty}{d\Omega}{\Omega^2 \hat{d}^*_J \hat{d}_k} P_{KJ}\, \, \,\,\,\ \mbox{for}\,\,\,\, K \neq J \,,
\label{worst}
\end{eqnarray}
where matrix  $\mathbf{P} =
\hat{\mathbf{A}}(\hat{\mathbf{A}}^\dagger\hat{\mathbf{A}})^{-1}\hat{\mathbf{A}}^\dagger$
($\hat{\mathbf{A}}$ is defined in Eq.~\ref{A_hat}).
Note that only $ J \neq K$ terms contribute in Eq.~\ref{T_geo}.
 
For the best-case scenario where the GW waveform is known and the only
unknowns are the initial wave arrival time $t_0$ and sky directions,
we found 
\begin{eqnarray}
\Delta_{KJ} &=& \frac{\xi_K \xi_J}{\sum_I \xi_I} \, \, \,\,\,\ \mbox{for}\,\,\,\, J \neq K
\label{best}
\end{eqnarray}
where we have defined
\begin{equation}
\label{xiJ}
\xi_J \equiv 2 \int_{-\infty}^{+\infty}\frac{d\Omega}{2\pi} {\Omega^2 |\hat{d}_J|^2}. 
\end{equation}
Note that $\xi_J$ corresponds to the noise-weighted GW energy flux coupled to the $J$th detector.

\subsection{Implication}
\label{imply}
Here we note the clear geometrical meaning of $ |(\mathbf{r}_{KJ} \times
\mathbf{r}_{ML})\cdot \hat{\mathbf{n}}|$ in Eq.~\ref{T_geo}, which is twice the area of the
quadrangle formed by the projections of detectors $J$, $K$, $L$
and $M$ onto the plane orthogonal to the wave propagation
direction.  We also note that, in the worst-case scenario where the
waveform is unknown, the angular resolution is inversely proportional
to the weighted correlation of responses between detectors (Eq.~\ref{worst}). In the best-case
scenario, it is inversely proportional to the fractional GW energy flux coupled to each detector (Eq.~\ref{best}, Eq.~\ref{xiJ}). 

Our formula is consistent with the known concept that a larger network is advantageous for a better
angular resolution.  For instance, inclusion of the future Australian AIGO detector can improve dramatically the angular resolution of the network \cite{wen07m}. It further indicates that angular resolutions can be
improved by optimizing values of $\Delta_{IJ}$.  For instance, building more detectors of correlated 
response is advantageous for localizing GWs of unknown waveforms.  Similarly, a specific localization method can be improved to approach the intrinsic angular resolution by selecting data contributing  significantly to 
fractional energy flux (best-case scenario) or to correlations of data between detectors (worst-case scenario). In other words, a localization method can be improved by proper treatments  of data corresponding to weak responses.

\section{Ranking Network Responses by Singular Value Decomposition Method}
In this and the next section, we demonstrate how one can improve the null-stream localization method for GWs of unknown waveforms. We show how to apply the singular value decomposition (SVD) method \cite{SVD} to recombine data from a network of GW detectors to form new data streams with characterized sensitivity to GWs from a sky direction.  Specifically, we use the SVD to construction signal streams, generalized null-streams that have  null responses to GWs, and semi-null streams  that have weak responses to signals.   The SVD of $\mathbf{\hat{A}}$ yields (see also \cite{wen07})
\begin{equation}
\mathbf{\hat{A}} = \mathbf{U} \mathbf{S} \mathbf{V}^\dagger,\mathbf{S}=\left(
                   \begin{array}{cc}
                     s_1 & 0 \\
                     0& s_{2} \\
                     0 & 0 \\
                         \vdots & \vdots \\
                            0 & 0 \\
                   \end{array}
                 \right),
\label{A_SVD}
\end{equation}
where $\mathbf{U}$ and $\mathbf{V}$ are unitary matrices of
dimensions of $N_d \times N_d$ and  $2\times 2$  respectively at each frequency, i.e., $\mathbf{U}\mathbf{U}^\dagger = \mathbf{I}$ and
$\mathbf{V}\mathbf{V}^\dagger =\mathbf{I}$, $s_1 \ge s_2\ge 0$ are the so-called singular
values. Note they are all frequency-dependent. 

We then construct new data streams by inserting this decomposition into equation $
\mathbf{\hat{d}} = \mathbf{\hat{A}} \mathbf{h}$, and have
\begin{equation}
\mathbf{U}^\dagger \mathbf{\hat{d}} = \left(
                   \begin{array}{c}
                     s_1 (V^\dagger \mathbf{h})_1\\
                     s_2 (V^\dagger \mathbf{h})_2\\
                     0\\
                         \vdots \\
                            0 \\
                   \end{array}
                 \right).
\end{equation}
It is evident that the first two components of the new data streams
contain signal information, the last
$N_d-2$ terms are null-streams as they have zero-response to signal. In general, null-streams can be written as 
\begin{equation}
{\mathbf{N}} (\Omega)=  \left(
                     \begin{array}{c} (\mathbf{U}^\dagger \mathbf{\hat{d}})_3\\
                                      (\mathbf{U}^\dagger \mathbf{\hat{d}})_4\\
                                       \vdots \\
                                      (\mathbf{U}^\dagger
                                      \mathbf{\hat{d}})_{N_d}\\
                      \end{array} \right).
\end{equation}
Naturally in case $s_2=0$, $(\mathbf{U}^\dagger \mathbf{\hat{d}})_2$-term should also be included as a null stream.  We assume for now that $s_2 \ne 0$.  The sensitivity level of each signal stream to a GW can be ranked by its singular value.   Suppose $s_{\mbox{max}} = \mbox{max}_{i,\Omega} s_i(\Omega)$ ($i=1,2$), we define (tentatively) semi-null streams as signal streams with corresponding singular values much less than  $s_{\mbox{max}}$. 
\begin{equation}
\mathbf{SN}(\Omega) = \left ( \begin{array}{c} (\mathbf{U}^\dagger \mathbf{\hat{d}})_{i}\\ \cdots\\   \end{array} \right )  \ \ \ \ \ \mbox{if}\ \ \ \ s_i \ll s_{\mbox{max}}\ \ \ \ i =1,2. 
\end{equation}
Although responses in semi-null streams may not be zero,  they can be insignificant compared to dominating signal streams. 

\section{Improved Localization Strategy Using Semi-Null Streams}
It has been demonstrated \cite{tinto89,wen05} that
null-streams ${\mathbf{N}} (\Omega)$ can be used to localize a source by searching through sky directions for minimum statistic of
\begin{equation}
P_N= \int d\Omega \sum_{I'} |{N}_{I^{'}} (\Omega)|^2.
\label{null}
\end{equation}
It has also been proposed \cite{wen07} that semi-null streams can be included to
improve the angular resolution.  One possible new statistic is 
\begin{equation}
P_{SN}= \int d\Omega \left (\sum_{I'} |{N}_{I^{'}} (\Omega)|^2 +
 \sum_{I'} |SN_{I'}(\Omega)|^2 \right ).
\label{semi_null}
\end{equation}
Instead of searching through sky directions for minimum statistic of
Eq.~\ref{null} as discussed in \cite{tinto89,wen05}, we now search for minimum statistic
in Eq.~\ref{semi_null}. The tricky part is how to set the threshold at which semi-null streams are to be included \cite{wen07b}.

\subsection{Numerical Example}

In this section, results from a Monte-Carlo simulation is presented to illustrate how the null-stream localization method can be improved by including the semi-null streams. As a proof-of-principle example, we have simply included all semi-null streams in Eq.~\ref{semi_null} that satisfied an 
empirical threshold of  $s_i(\Omega)/s_{\mbox{max}} \le
0.01$ (where $s_i$ are singular values defined in Eq.~\ref{A_SVD}). Localization
is then obtained by searching through sky directions for minimum statistic of 
(1) null-stream-only statistic (Eq.~\ref{null}) 
and (2) semi-null-stream statistic (Eq.~\ref{semi_null}) respectively.  Results are then compared. 

We have used simulated signal and noise.  For the signal, we used a
Sin-Gaussian wave form of $ h_+(t)=h_\times(t)=h_0 \sin(2\pi
f_0t)\exp(-t^2/\tau^2)$ with polarization angle chosen arbitrarily to
be $ \psi=0$, signal duration $T=7$ ms, sampling rate $= 16$ kHz, central
frequency $f_0=700$ Hz and $\tau=2$ ms. The arrival time of the GW
wavefront at LIGO Livingston (L1) is chosen arbitrarily to be at 0.00
hr, March, 18, 2004.  The source direction was chosen to be near that of
the maximum sensitivity of L1 (right ascension RA$=85.1235^0$ and
declination Dec$= 30.56^0$) at the chosen time. We have chosen an 
optimal network signal-to-noise ratio of SNR$=20$.  Location
information of different GW observatories were obtained
from \cite{allen96} and references therein.  For the noise, we have
adopted the designed noise spectral densities for initial LIGOs (at
Livingston, L1, and at Hanford, H1) \cite{ligonoise} and for
GEO \cite{geonoise} at $500$ Hz tuning.  The simulated GW signal is
then injected into a total of 500 sets of randomly generated Gaussian
noise.  For each of the simulated data of noise plus
signal, we use the Nelder-Mead method \cite{recipe} to search through sky directions
for minimum statistics of Eq.~\ref{null} and
Eq.~\ref{semi_null} respectively. All searches start from the source
direction to shorten the search time which is adequate for the purpose of proof of principle. 

\begin{figure}
\begin{center}
\includegraphics [scale=0.32]{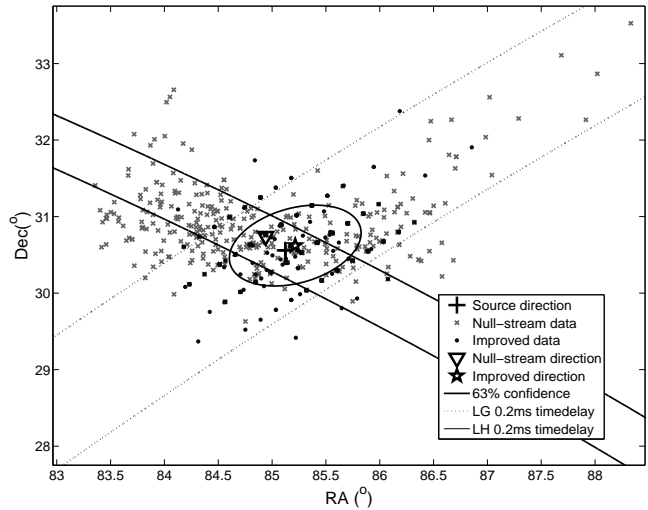}
%\vspace*{0.2in}
 \caption{Comparison of localization methods. }\label{ff}
\end{center}
\end{figure}

The results are shown in Fig.~\ref{ff}.  Source directions
obtained using the null-stream-only method (Eq.~\ref{null}) for different noise realizations are plotted in cross symbols. Filled circles are those from the improved localization method where semi-null
streams are included (Eq.~\ref{semi_null}).   The error
ellipse is for data from the improved method at a $63$\% 
confidence level assuming a  bi-variate normal distribution of angular parameters.  The star symbol indicate the average direction in the improved
method.  The actual source direction is
indicated with a cross.  The triangle symbol indicates the average direction from the null-stream-only 
method.  We also plot the time-delay lines for L1-G (gray dotted lines) and
L1-H1 (solid lines) at a 0.2 ms interval.  

It is evident that in this
particular example, for a direction where sources are most likely to be detected by
the LHV-network, inclusion of semi-null streams can improve the source localization significantly.  The scatter
of angular directions obtained from the null-stream-only (gray crosses)
method is much larger (and therefore worse localization) than that when semi-null
streams are also included (filled circles).  Note that our choice of the semi-null stream is not optimal, further
improvements are expected when optimal search methods are constructed \cite{wen07b}.

 \section{Conclusion}\label{conclusion}
We have reported for the first time a compact geometrical expression
for an arbitrary network of GW detectors when the initial arrival time
of a GW is unknown.  Our results demonstrate the known geometrical elements, as well as the role of energy fluxes and correlation between responses of different detectors  that determine the intrinsic angular resolution of a GW detector array.  In the second part of this paper, we show an example where localization of a GW source can be improved by including semi-null streams which are linear combination of data that have
weak response to a GW signal than that of null-stream-only method.  We show how the Singular-Value-Decomposition method, besides providing a vehicle for generalized optimizations of detection methods and for construction of generalized null-streams \cite{wen07}, can also be used to identify semi-null-streams and help improve the GW source direction determination. 

\ack

This work is supported by the Alexander von Humboldt Foundation's Sofja Kovalevskaja Programme funded by the German Federal Ministry of Education and Research. YC is supported in part by the NSF grants PHY-0653653 and PHY-0601459, and by David and Barbara Groce start-up fund at Caltech.  XF is supported in part by the National Natural Science Foundation of China, under Grant No. 10533010, 973 Program No. 2007CB815401 and Program for New Century Excellent Talents in University (NCET) of China.

\end{document}